\documentstyle[prb,aps,epsf]{revtex}

\draft
\title{On Effective  Electron Mass of Silicon MOSFET at Low Electron Density.}
\author{V.T.~Dolgopolov}
\address{Institute of Solid State Physics, RAS,
Chernogolovka,142432 Russia}

\begin{document}
\maketitle

\begin{abstract}
 The trial wave function method developed in Ref.s \cite{gutz,brink}
for the case of narrow {\it s}-band in a perfect crystal is adapted for calculation of the density dependence of the effective mass and the Lande factor in a dilute two-dimensional electron system. We find that the effective mass has a tendency to diverge at a certain critical concentration, whereas the $g$ factor remains finite.
\end{abstract}

As the temperature decreases, a dilute electron gas  of highly
mobile Si-MOSFET exhibits a strong drope in the resistance if the
electron density $n_c$ is higher than a certain critical one
$n_s>n_c$, and an inrease in resistance when   $n_s<n_c$
\cite{krrev}. In the vicinity of  $n_c$  the resistance possesses
scaling properties as a function of temperature and  electron
density.  This instance from the very first caused researchers to
consider the observed transition as a disorder-controlled quantum
phase  metal-insulator transition (MIT) and gave rise to a tide of
similar investigations of other systems in which any change in the
sign of derivative  $\frac{dR}{dT}(n_s)$ was taken as an evidence
of the occurrence of a quantum phase MIT.

In the recent experimental study \cite{shash} of the screening
properties of a two-dimensional electron system as a function of
temperature \cite{gold}, the interpretation was used in the terms
of Ref. \cite{ZNA} , and a strong increase of the effective mass
was found in Si-MOSFET as the electron density approached  a value
of $\sim 0.8*10^{11}$ cm$^{-2}$,  which is close to $n_c$ in the
best of investigated samples. A similar behavior of the cyclotron
mass was observed in independent  \cite{pud} experiments on the
measurement of the temperature dependence of Shubnikov-de Haas
oscillations. An analysis of the experimental data similar to that
made in \cite{shash} but performed in the opposite limit in the
ratio of valley-splitting energy  to temperature with the use of
data of other experimental groups and samples from other sources
\cite{vitk,pud1}, confirmed the versatility of the $m^*(n_s)$
curve.

   The conclusion that should be made from recent experimental data is that the quantum phase transition observed in the most perfect MOSFET is rather the property of a pure disorder-free two-dimensional system. A qualitative theory of two-dimensional electron Fermi liquid in a state close to crystallization was presented in  Ref.s \cite{spivak,spivak1}.
Below a quantitative description of a  two-dimensional paramagnetic electron liquid is proposed, adapting the trial wave function approach developed in Ref.s
\cite {gutz,brink}for the case of  a narrow $s$ band in a perfect crystal.

Let us assume that the ground state of an electron system with
strong interaction in a regime close to crystallization can be
described as an crystal with a great number of charge -carrying
mobile defects. The real two-dimensional system will be replaced
by a grid of lattice sites with a density of $n_s$. An electronic
wave function of the Wannier type $\phi[({\bf r-g})n_s^{1/2}]$ ,
where the vector  ${\bf g}$  specifies the position of a lattice
site, will be associated with each site. The corresponding
creation operator is $a^{\dagger}_{\bf g}$. If each site were
occupied by only one electron, the system would represent a
perfect electron crystal. In fact, there is a certain probability
depending on $n_s$ that in the ground state a site can be occupied
by two electrons with opposite spins. The number of such states
eventually determines the number of  mobile excitations  and,
hence, the transport properties of  the system.

Bloch wave functions are constructed on the base of lattice sites:
\begin{equation}
\psi_{\bf k}({\bf r})= n_s^{-1/2} \sum_{\bf g}exp(i{\bf k}{\bf
g})\phi({\bf r-g}), \label{eq1}
\end{equation}

\begin{equation}
 a^{\dagger}_{\bf k}= n_s^{-1/2} \sum_{\bf g}exp(i{\bf k}{\bf
g})a^{\dagger}_{\bf g}. \label{eq2}
\end{equation}
The Hamiltonian of the system contains the electron kinetic energy
and the electron interaction at one site:
\begin{equation}
{H}= \sum_{\bf k}\varepsilon_{\bf k}(a^{\dagger}_{{\bf
k}\uparrow}a_{{\bf k}\uparrow}+a^{\dagger}_{{\bf
k}\downarrow}a_{{\bf k}\downarrow})+\frac{\alpha
e^2}{\varepsilon_0 n_s^{1/2}} \sum_{\bf g}a^{\dagger}_{{\bf
g}\uparrow}a^{\dagger}_{{\bf g}\downarrow}a_{{\bf
g}\downarrow}a_{{\bf g}\uparrow}.
 \label{eq3}
\end{equation}
Here,  $\varepsilon_{\bf k}=\frac{\hbar^2 k^2}{2m}$,  and
$\varepsilon_0$  is the static dielectric constant. We introduced a coefficient
 $\alpha$ into the interaction energy determined by the exact form of the wave function on the site  and completely  neglected the electron interaction on the neighboring sites.
Processing to the limit of the gas of noninteracting electrons
requires the modulation in Eg.(1) disappear and the coefficient
$\alpha$   be a slowly varying function of $n_s$, vanishing at
$n_s^{-1} \rightarrow 0$. We will neglect this weak dependence in
the region of a low electron densities.

In Ref.~ \cite{gutz}, it was proposed that a many-body trial function be used for a ground state in the form

\begin{equation}
\psi=\sum_{G\Gamma}A_{G\Gamma}\prod_G a^{\dagger}_{{\bf
g}\uparrow}\prod_{\Gamma}a^{\dagger}_{{\bf g}\downarrow}\Phi_0,
\label{eq4}
\end{equation}
where $G$  and  $\Gamma$  are the sets of sites occupied by electrons with spin up
and down, repectivelly; and $\Phi_0$ is a vacuum state. It is convenient  to express  the function $\psi$ through operators of creation and annihilatin of Bloch waves and to take into account electron correlation by decreasing coefficient
$A_{G\Gamma}$ in  (\ref{eq4})
by a factor of  $\eta^\nu$ if the corresponding product implies the occurence of doubly occupied sites whose  fraction  equals  $\nu$ ($0<\eta<1$).
The relation between $\nu$ and  $\eta$  in the ground state was obtained  in \cite{gutz}.
For our case,
\begin{equation}
\eta=\nu \Big(\frac{1}{2 } -\nu \Big)^{-1}. \label{eq5}
\end{equation}
The probability that a single-particle state with a wave vector
${\bf k}$  is occupied undergoes a jump at $k=k_F$  by the value
\begin{equation}
q=16\nu \Big(\frac{1}{2 } -\nu\Big). \label{eq6}
\end{equation}
Thus, the trial wave function describes a mixture of functions that corresponds to a fully occupied  band (solid spin-ordered phase) and a paramagnetic  electron liquid. The transition to the solid phase is continuous  and is characterized by the parameter
$q$ ($0\leq q\leq 1$): $q=1$ in the paramagnetic electron liquid with weak interaction, and  $q=0$ in the electron crystal.

The mean value of the Hamiltonian given by   (\ref{eq3}) in the state with given  $\nu$ equals
\begin{equation}
\langle
H\rangle_{\nu}=\frac{1}{2}n_sq\varepsilon_F+\nu\frac{\alpha
e^2}{\varepsilon_0}n^{3/2}, \label{eq7}
\end{equation}
where   $\varepsilon_F$  is the Fermi energy   of an equivalent number of electrons in the absence of interaction. According to  \cite{brink}, the expression in Eq. (\ref{eq7})
is minimized with respect to  $\nu$  with regard to Eq.(\ref{eq6}).
A minimun of the Hamiltonian is attained at
 \begin{equation}
\nu=\frac{1}{4}\Big(1-(\frac{n_{c1}}{n_s})^{1/2}\Big);
n_c1=\Big(\frac{\alpha e^2m}{2 \varepsilon_0\pi\hbar^2}\Big)^2;
\label{eq8}
\end{equation}
which,according to (\ref{eq6}),  corresponds to
\begin{equation}
q^{-1}=\frac{m*}{m}= \frac{n_s}{n_s-n_{c1}}. \label{eq9}
\end{equation}
Here,  $m*$   is the renormalized effective mass. In the same way, following  \cite{brink},
the  Lande factor can be found as
\begin{equation}
\frac{g*}{g}=\Bigg[1-(\frac{n_{c1}}{n_s})^{1/2}\frac{1+\frac{1}{2}(\frac{n_{c1}}{n_s})^{1/2}}{\big(1+(\frac{n_{c1}}{n_s}
)^{1/2}\big)^2}\Bigg]^{-1}, \label{eq10}
\end{equation}
The simplest  of way  of generalization to the case of two valleys  is in considering  two parallel sublattices. In each of them  the number of electrons  equals
$n_s/2$ and the characteristic cell size is deminished compared to the single-valley case by a factor of  $\beta$. The coefficient $\beta$
is determined by the ratio of Coulomb energies of inter and intravalley interactions. In the limit of two sublattices in one plane  $\beta=\sqrt2$ .
In the case of two valleys, $n_{c1}$  in  Eq.~(9,10) should be replaced by
$n_{c2}=2\beta^2 n_{c1}$.

A comparison of the curves obtained in this way with experimental results is shown in Figs.~ 1,2. Single fitting parameter   $n_{c2}=0.78*10^{11}$ cm~$^{-2}$  has been used, which corresponds to $\alpha=0.15$.  It is evident from the figures that the behavior of both the effective mass and the   $g$- factor is reasonably described within framework of the proposed model, though  the coefficient  $\alpha$  is approximately twice as large as the value expected
 according to numerical calculations \cite{tana}.

It should be specially noted that the above considerations give no
way of judging  the spin state of the solid phase, because it is
determined by  the exchange interaction of electrons on
neighboring sites.  Moreover, in the immediate vicinity of the
transition point, in the region where  $(\langle
H(n_s)\rangle_\nu-\langle H(n_c)\rangle_\nu)n_s^{-1}$ turns out to
be smaller than the exchange energy of electrons on neighboring
sites, the proposed description does not work in the paramagnetic
electron liquid as well. Thus, the issues of the phase diagram in
the immediate vicinity of the transition point and those of the
spin structure of the solid phase remain out of scope of this
consideration. In general, the approximation used is poorly
controlled, and the rather good description of experiment still
remains its only justification.

It is well known that the concentration  $n_c$ that corresponds to a change of the sign of the derivative   $dR/dT$  strongly varies from sample to sample, depending on the disorder  in the electron system under study. An impression is gained from  experimental data that the transition point  $n_{c2}$ measured for different samples  also somewhat varies. This fact can also be related to the effect of disorder, as was discussed in  \cite{spivak,spivak1}.
In strongly disordered electron systems $n_c\gg n_{c2}$  and the effect considered above, namely, the dramatic increase in the effective mass is not observed. In the most perfect of the electron systems studied, $n_c\approx n_{c2}$.

Author is grateful to V.F.~Gantmakher, A.~Gold, S.V.~Iordanski, B.~Spivak, D.E. Khmelnitskii, and A.A.~Shashkin for useful discussions. A significant part of this work was carried out at the LMU, and the author is grateful to J.P.~Kotthaus and researchers from his institute for help and discussions.

This work is supported by the Russian Foundation for Basic Research, the Ministry for Science and Technology of the Russian  Federation, and by A.~von Humboldt  Forschungspreis.

\begin{figure}[t]
\centerline{ \epsfxsize=\columnwidth \epsffile{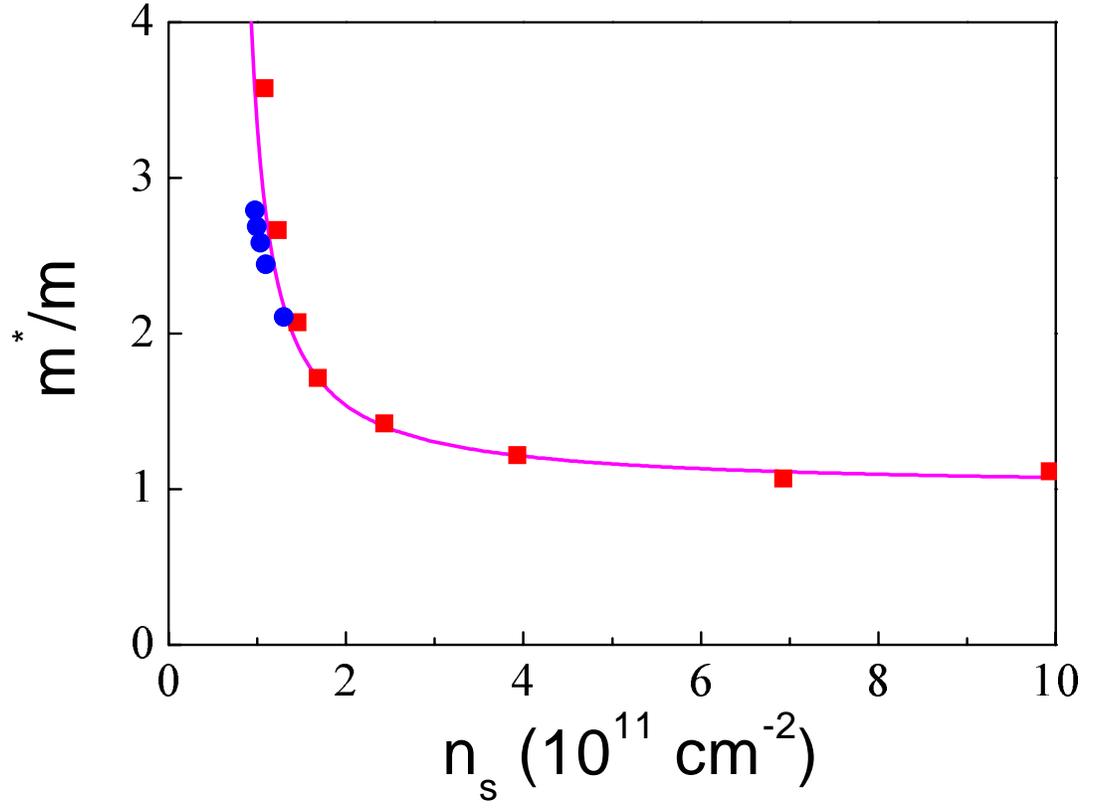} }
\caption{ Effective mass as a function of electron density. The
solid line corresponds to Eq.~\protect\ref{eq9} with
$n_{c2}=0.78*10^{11}$ cm~$^{-2}$. Squares and circles correspond
to the experimental data from  Ref.~\protect\cite{shash},  and
Ref.~\protect\cite{pud2}, respectively.} \label{fig1}
\end{figure}

\begin{figure}[t]
\centerline{ \epsfxsize=\columnwidth \epsffile{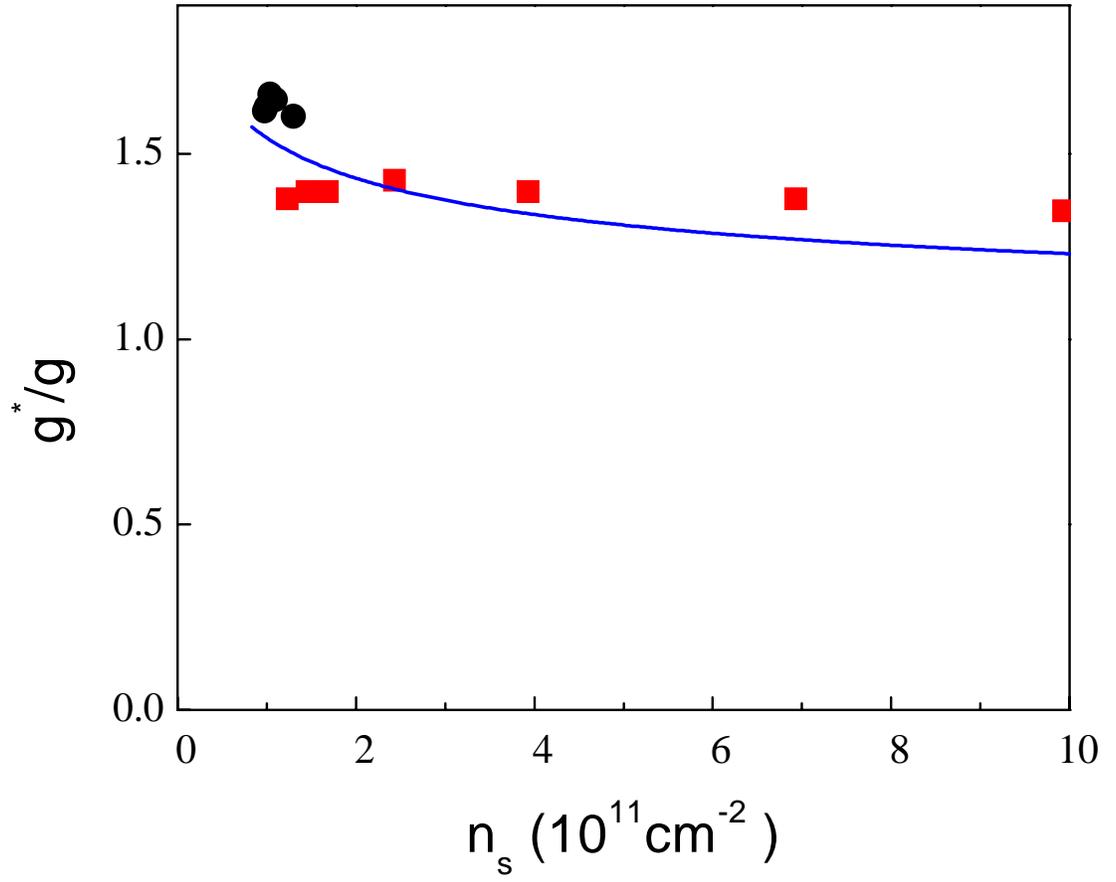} }
\caption{Effective  $g$- factor  as a function of electron
density. The designations of experimental points are the same as
in  Fig.~\protect\ref{fig1}.} \label{fig2}
\end{figure}

\end{document}